# Characterization of 3D Printers and X-Ray Computerized Tomography

## Sunita Khod[1], Akshay Dvivedi[2], Mayank Goswami[1,#]

*[1]Divyadrishti Imaging Laboratory, Department of Physics,* Indian Institute of Technology Roorkee, Roorkee, Uttarakhand-247667, India.

*[2]Department of Mechanical and Industrial Engineering,* Indian Institute of Technology Roorkee, Roorkee, Uttarakhand-247667, India

[#]mayank.goswami@ph.iitr.ac.in

## Abstract

The 3D printing process flow requires several inputs for the best printing quality. These settings may vary from sample to sample, printer to printer, and depend upon users' previous experience.

The involved operational parameters for 3D Printing are varied to test the optimality. Thirty-eight samples are printed using four commercially available 3D printers, namely: (a) Ultimaker 2 Extended+, (b) Delta Wasp, (c) Raise E2, and (d) ProJet MJP.

The sample profiles contain uniform and non-uniform distribution of the assorted size of cubes and spheres with a known amount of porosity.

These samples are scanned using X-Ray Computed Tomography system. Functional Imaging analysis is performed using AI-based segmentation codes to (a) characterize these 3D printers and (b) find respective optimal input settings.

Three-dimensional surface roughness of three teeth and one sandstone pebble (from riverbed) with naturally deposited layers is also compared with printed sample values. Teeth has best quality.

It is found that ProJet MJP gives the best quality of printed samples with the least amount of surface roughness and almost near to the actual porosity value. As expected, 100% infill density value, best spatial resolution for printing or Layer height, and minimum nozzle speed give the best quality of 3D printing.

Key Words: 3D Printing, X-Ray Computed Tomography, Cusp Density, Surface Roughness.

## 1. Introduction

The field of 3D Printing is expanding at a faster rate. Additive Fabrication, Additive Process, Additive Manufacturing, and Additive Technique are the other synonymous used for 3D Printing [1]. From the manufacturing of toys to the prototyping of automobiles, 3D printing is used in every domain, in one way or the other. The technology of 3D Printing is also used today to produce functional metal parts for applications in medical and aerospace. 3D printed Radiation Shields are used in Nuclear for protection from nuclear radiation [2]. Apart from it, there are numerous other applications as well such as in the field of building construction and the food industry. The ability to produce complex parts, less material wastage and less time consumption make the process of 3D Printing more widely used today.

According to the standards defined by the American Society for Testing and Materials, the process in which most 3D Printers work is classified into seven categories which include Vat Polymerization, Material Jetting, Binder Jetting, Material Extrusion, Powder Bed Fusion, Direct Energy Deposition, and Sheet Lamination. The process of 3D Printing can be understood as the interaction between the mass of the material and the energy required to form a single layer. The process of Powder Bed Fusion, Binder Jetting, Sheet Lamination, and Vat Polymerization falls in the category of a 3D Printing process in which material mass is variable.

Material Extrusion and Material Jetting are grouped as variable energy processes. In Direct Energy Deposition, both material mass and energy are variable while forming a single layer [3].

Fused Deposition Modeling (FDM) is an example of the Material Extrusion Process. In the process of FDM, the material is filled in the nozzle in the form of a filament, heated, and melted material is deposited layer by layer as the nozzle of the printer moves and part is fabricated [4]. However, the part fabricated by the process of Fused Deposition Modeling is not accurate and differs from the ideal geometry due to the manufacturing process. Surface roughness and porosity are common defects present in the 3D printed part. It is expected that the location of failure coincides with the location of these defects. Therefore, to check the quality of the part printed, an inspection of the part fabricated using the 3D Printing process is necessary, especially for high value and critical parts used for aerospace and medical applications [5].

Computed Tomography (CT) is a technique for the 3D representation of an object. X-Ray CT is a non-destructive technique for the 3D assessment of the internal defects present in the material [6]. X-Ray CT systems are used in medical and industrial applications. The resolution of industrial or Micro-CT is in the range of micrometer scale. With the advancement in technology, resolution in the range of nanometer is also achievable today.

3D Printing and X-Ray micro-CT thus are complementary technology these days and has been growing rapidly. With the increasing demand for 3D printed parts and samples, it is important to test the quality of the samples and parts printed. This can be non-destructively done by the X-Ray micro-CT. Analyzing the 3D Printing technology using X-Ray CT is one of the industrial applications of X-Ray micro-CT. The main applications of X-Ray CT in AM are defect detection, dimensional evaluation, density measurement, and surface roughness analysis [7].

## 1.1 Motivation

The 3D printed sample may have a deviation from the actual Computer-Aided Design (CAD). That means there is some error in the accuracy of the 3D Printer. The resolution and printing quality of any 3D printer may differ from the technical specifications given by the manufacturer as its operational age approaches. This study aims to compare the quality of the different commercially available 3D Printers using micro–X-ray CT. Measurement of surface roughness of samples printed from 3D Printers is one of the methods to check the quality of the printed samples. Surface roughness is the critical parameter in determining the surface texture of the printed objects [8]. Different printing parameters such as printing temperature, nozzle diameter, layer height, speed of the nozzle, extrusion speed, infill density, infill pattern, bed temperature, and material properties govern the roughness in the printed part.

The high value of printing temperature and speed may lead to more surface roughness. Similarly, higher layer height may lead to coarser quality print [9]. One can set extreme / best printing settings for a smoother / finer finish, which in the first place may be overkill or not exactly desired or consume a relatively lot of resources and printing time. Therefore, an optimal setting of the printer may vary depending upon the required texture/quality.

## 2. Theory

### 2.1 X-Ray Computed Tomography

The three main components of the X-Ray Computed Tomography system are the X-Ray tube, object, and detector. The sample is irradiated with X-Rays from the X-Ray tube. The X-Rays get attenuated after passing through the sample. The attenuation of X-Rays follows the Lambert-Beer Law which states that the attenuation of the X-Ray beam occurs exponentially. Mathematically it is given by: -

$$I = I_0 * e^{-\mu t}$$

Where I and $I_0$ are the intensity of the X-Ray beam after and before passing through the sample, $\mu$ is the Linear Attenuation Coefficient and t is the thickness of the sample [10]. The attenuated intensity of X-Rays is collected by the detector in the form of projection data as the object rotates through 360º. The X-Ray CT system used for this study is shown in Figure 1.

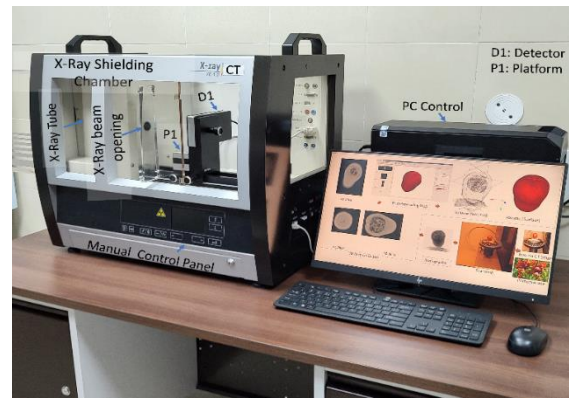

Figure 1: X-Ray Computed Tomography

The acquired radiographs or 2D projection images are reconstructed to form slices and volumetric data. An integrated software based on reconstruction algorithms is required to reconstruct the data. Filtered Back Projection, Feld Kamp Davis Kress, Hilbert Filtered Back Projection, and iterative reconstruction are some of the reconstruction algorithms most widely used. After the reconstruction, data is analyzed for further visualization.

The quality of visualization of CT scan data depends upon the resolution of the X-Ray CT system. The resolution of a CT system in turn depends upon the energy of the X-Ray beam used. There are systems available that utilize soft X-Rays having an energy of few keV and a resolution of the order of microns. In CT systems utilizing hard X-Rays of energy of 100 keV and more, a resolution of 1 nm is achievable.

### 2.2 3D printing Process

3D printing is the process of making 3D objects from a CAD model. The object which needs to be 3D printed is the first modeled using CAD software. Solidworks, AutoCAD, CATIA, and ARCHICAD are some of the commercially available software used for CAD modeling. After CAD modeling, the file is converted to .STL files. STL is the extension used for Standard Triangle Language or Standard Tessellation Language files. The .STL file is loaded into slicing software such as Cura and Idea maker. The slicing of the sample is done and printing parameters are set and converted to .gcode files. The .gcode files are then loaded into the 3D printer. For the additive manufacturing systems utilizing the process of FDM, the material is filled in the nozzle of the printer in the form of the filament. The material is then heated inside the nozzle. The nozzle temperature varies from 220º C to 240º C. The nozzle of the printer moves under the computer control and extrudes and deposits the melted material layer by layer on the bed. The bed temperature varies from 30º C to 55º C. The speed of the nozzle varies from 30 mm/s to 150 mm/s. The nozzle deposit the material till the whole sample is printed. The printer stops after the printing of the sample finishes. The printed part needs to be post-processed to remove the extra part which is added as a base while printing the 3D sample.

The printing parameters such as printing speed, printing temperature, layer height, infill density, and infill pattern need to be adjusted in slicing software before printing. The printer may fail to print the required sample if the parameters are not according to the requirement of printing. Therefore, printing parameters need to be adjusted such that the part gets printed with fine quality. This setting of the printer for which the sample gets printed with good resolution is called the optimal setting of the printer.

### 2.3. Roughness Measurement and Cusp Density Estimation

A cusp is a point of transition or a pointed end where two curves meet. Cusp density is defined as the ratio of cusp volume to the total volume of the sample. Minimum cusp density means smoother the surface of the sample. Roughness is measured as a deviation in the geometry of the printed sample from the actual dimensions of CAD. Cusp Density

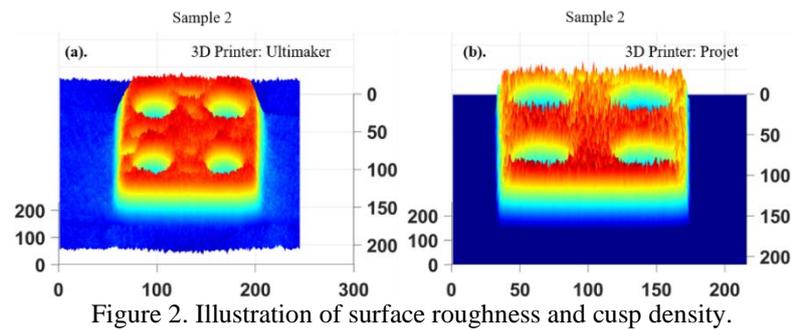

Figure 2. Illustration of surface roughness and cusp density.

and roughness measurement analysis is done to estimate the surface roughness of 3D printed samples.

Figure 2 illustrates the difference between 3D printed samples qualitatively. Fig. 2(a) and Fig. 2(b) show the cusp surface of an X-Ray CT Slice (at a particular height) of a 3D printed sample with maximum and minimum cusp distribution. MATLAB Code to estimate these parameters is provided as a supplementary file.

## 3. Methodology

To check the quality and optimal setting of different 3D Printers, we printed two samples having different dimensions from Ultimaker Extended 2+, Delta Wasp, Raise E2, and ProJet for six different printer settings. The material used for printing all the samples is Polylactic Acid (PLA) which is a biodegradable thermoplastic polyester. The samples are printed at different layer heights and speeds. The bed temperature, nozzle diameter, infill density, and infill pattern are kept the same for all the printers and all the samples. For quality testing, all the samples are scanned using an X-ray CT system, and the internal and outer structure is analyzed. The surface roughness for both the samples at all the settings is analyzed. The optimal setting of different 3D printers is found.

### 3.1 Experimental Procedure

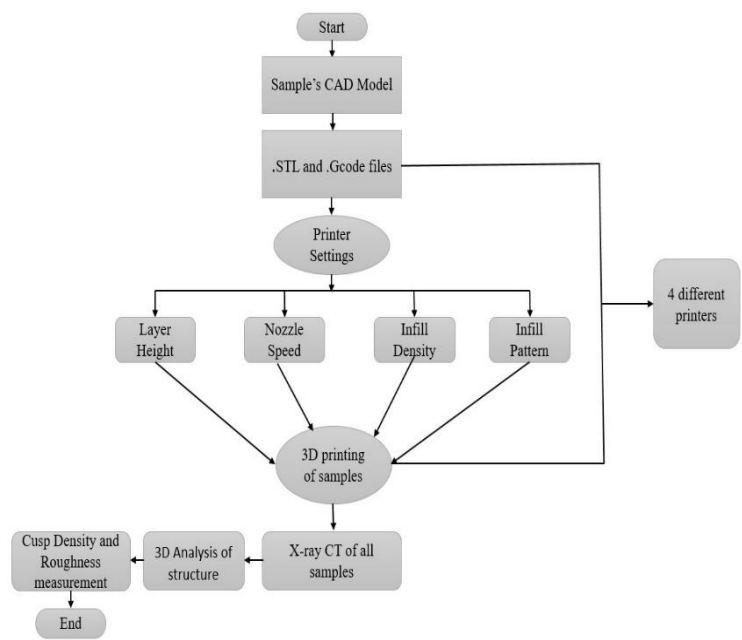

Figure 3: Flowchart for Adopted Methodology

The process flow is shown in Figure 3. The adopted methodology for the characterization of additive manufacturing systems using an X-ray CT system consists of five steps. Firstly, the samples are modeled using CAD software. To check the optimality of the printer, the samples are printed with various printer settings from four different types of 3D printers. The printed samples are then CT scanned using an X-ray CT system and a 3D analysis of data is done for the characterization. Surface roughness is determined on the basis of cusp density and roughness parameters.

### 3.2 CAD of samples

To determine the quality of the different 3D Printers, two samples are modeled in Solidworks. The dimensions of the first and second samples are 8mm (L) × 8mm (W) × 26mm (H) and 6mm (L) × 6mm (W) × 17.50mm (H) respectively. The first sample has cuboid shaped cavities of variable dimensions (top to bottom) are inserted inside the sample. The dimension of the largest and smallest void are 1.40mm (L) ×1.40mm (W) × 1.40mm (H) and 0.20mm (L) × 0.20mm (W) × 0.20mm (H). The dimensions of other cubes present inside the first sample are in between the largest and smallest cube's dimensions. The pattern followed for inserting the voids is such that cuboids of large and small dimensions are inserted alternatively. The dimension of large cubes decreased from top to bottom while the dimensions of the small cube first increased from top to middle of the sample and then decreased to the bottom. The second sample is inserted with cavities/voids in the shape of spheres and cubes having variable dimensions. The diameter of the largest and smallest spheres is 1.20mm and 0.20mm. The dimensions of the largest and smallest cubes are 0.70mm (L) × 0.70mm (W) × 0.70mm (H) and 0.20mm (L) × 0.20mm (W) × 0.20mm (H). The dimensions of other voids are in between the dimensions of the largest and smallest voids. The spheres and cubes are inserted alternatively and the dimensions of both types of voids decreased from the top to the bottom of the sample such that the largest dimension void lies at the top and the smallest dimension void lies at the bottom as shown in Figure 4 below.

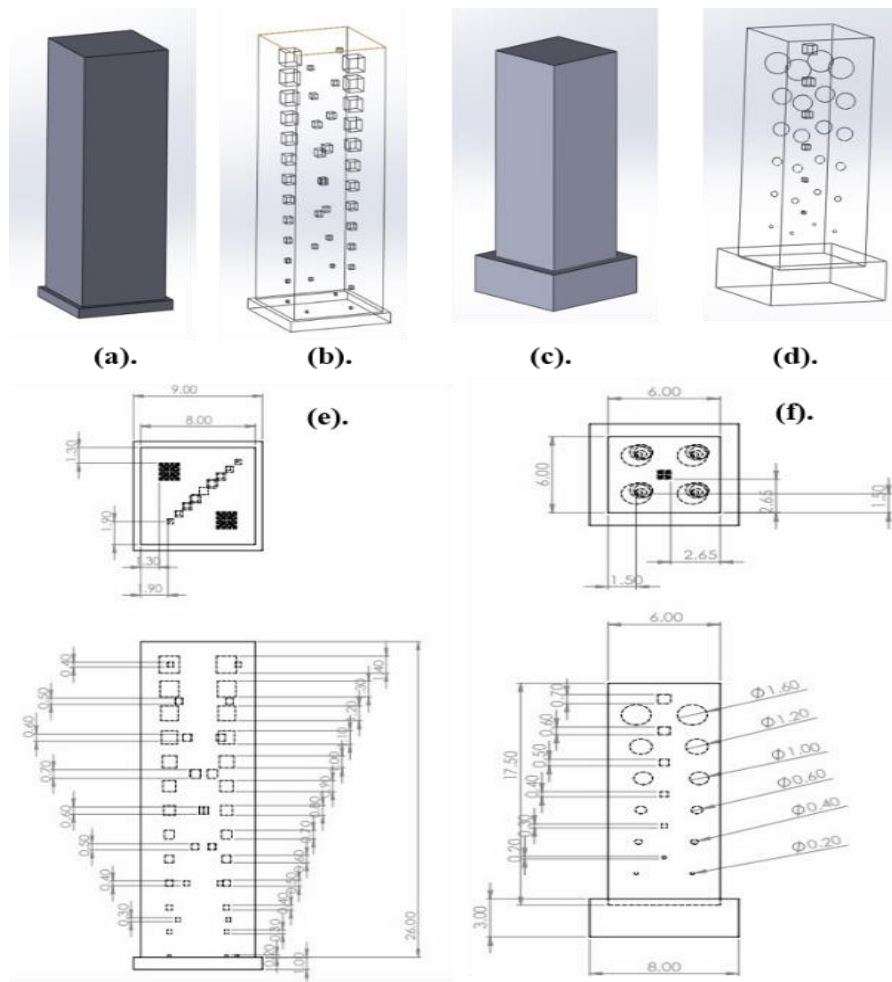

Figure 4: CAD showing Void distribution: (a). CAD image of Sample 1, (b). Voids in Sample 1, (c). CAD image of Sample 2, (d). Voids in Sample 2, (e). Size distribution of voids in Sample 1, (f). Size distribution of voids in Sample 2.

### 3.3. Printer Type and Material Used

The CAD models of both the samples are 3D printed for all the six printer settings listed in Table 1 and from four different 3D printers. The four 3D printers used are Ultimaker Extended 2+, Delta Wasp, Raise E2, and ProJet MJP. The resolution of Ultimaker Extended 2+ varies from 200 microns to 20 microns for a 0.40 mm nozzle diameter. The minimum resolution for Delta Wasp is 50 microns. The ProJet MJP 3600 is an industrial printer that works in HD, UHD, and XHD modes. The resolution of ProJet in HD, UHD, and XHD modes are 32 microns, 29 microns, and 16 microns respectively. The minimum resolution with which Raise E2 works is 50 microns. The material used for printing the samples is Poly-lactic Acid (PLA). The economical production of PLA from renewable resources and its biodegradable nature makes it the most widely used plastic filament material in 3D Printing.

### 3.4 Printer Settings

The quality of the sample printed from a 3D Printer depends upon various settings. Roughness parameter as a function of printer settings is estimated as quality criterion. Several experiments are designed to print the samples at different printer settings. The layer height or resolution is varied from 50μm to 70μm while keeping nozzle speed 30mm/s, and infill density 100%. The infill pattern is Grid. The Nozzle Temperature is kept at 220°C and the Nozzle diameter is 0.4mm. The bed temperature of the printer is kept at 45° C. The samples are then 3D printed for the given list of printer settings. Different printer settings print the samples with different amounts of surface roughness. The optimality of the printer setting for both the samples and for all the four printers is checked. The list of the printer setting is given in Table 1.

**Table 1:- List of Printer Settings**

| S. No. | Layer Height (μm) | Nozzle Speed (mm/s) | Infill Density (%) | Infill Pattern |
|--------|-------------------|---------------------|--------------------|----------------|
| 1.     | 50                | 30                  | 100                | Grid           |
| 2.     | 55                | 30                  | 100                | Grid           |
| 3.     | 60                | 30                  | 100                | Grid           |
| 4.     | 65                | 30                  | 100                | Grid           |
| 5.     | 70                | 30                  | 100                | Grid           |
| 6.     | 50                | 35                  | 100                | Grid           |

### 3.5. 3D analysis of data

All the samples printed from four different 3D Printers at different printer settings are scanned using an X-Ray CT system with 35 kV and 1mA applied across the X-Ray tube. The data is obtained in the form of projections. The reconstructed slices are formed from the projection data using integrated software based on the Filtered Back Projection reconstruction algorithm. The 3D analysis and volume rendering analysis of the structure of the samples are done by in-house developed codes written in MATLAB. The surface roughness is determined for all the samples using AI-based segmentation code. The printer setting and the type of 3D Printer with minimum surface roughness are found.

## 4. Experimental Results and Discussion

### 4.1. 3D Printed Samples

In total, 38 samples are 3D printed. This number is due to 6 printer settings for 2 samples of different dimensions, by three printers: Ultimaker, Delta wasp, and Raise E2. From ProJet MJP, one copy of the CAD model of Sample 1 and Sample 2 at 16-micron resolution is printed. The time taken (2 hr 30 min) for printing the samples is approximately same for all the 3D printers. The samples printed from different printers at different printer settings are shown in Figure 5. The samples with the texture of blue color are printed from Delta Wasp, with black color from Ultimaker, white from Raise E2, and showing transparent texture from ProJet MJP.

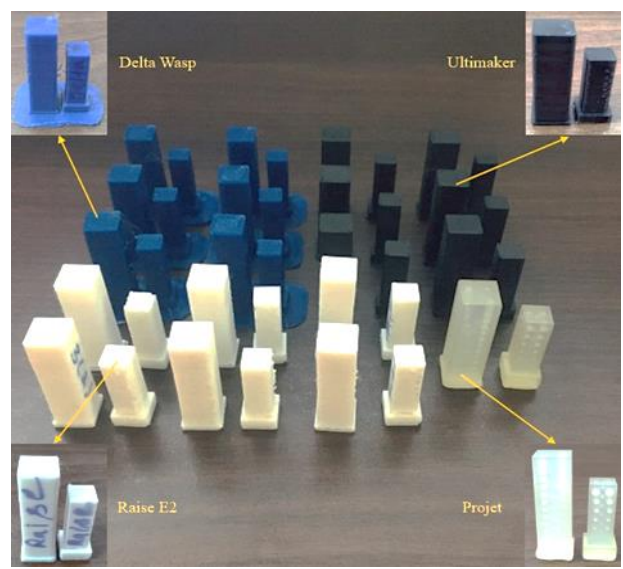

Figure 5: 3D Printed Samples:

## 4.2. Volumetric and X-Ray CT slice Analysis

The visualization of the 3D volume of all the samples fabricated from different 3D Printers is done using codes written in MATLAB. The volume rendering created the internal volume and void size distribution of the sample structure. It is observed that for both the samples printed from Delta Wasp, Raise E2, and Ultimaker, voids of size 0.3mm (with 50% accuracy) and above are printed and are visualized in the volumetric analysis. For samples 1 and 2 printed from ProJet, voids of size 0.2mm (with 100% accuracy) and above are printed by the process of 3D Printing.

X-Ray CT slice at a certain height in the volume rendering of the sample is also shown. It is observed that the voids are not printed in sample 1 by the Delta Wasp 3D printer in correct/perfect shapes. The voids printed by Raise are close to the size of the inserted void. The voids are printed with maximal distortion by Ultimaker Extended 2+. The internal volume of sample 1 printed by Ultimaker is distorted and defects are present. This may be due to the reason that either the performance of the printer is decreased, or the nozzle speed is not accurate for printing the voids perfectly. ProJet is able to print the perfect-sized voids. The same conclusion holds true for sample 2. It is also observed that the surface of cubical voids in sample 1 is rough as compared to the surface of spherical voids in sample 2, irrespective of the type of 3D printer used. From the volumetric and X-Ray CT slice analysis, it is concluded that the resolution of ProJet is fine and prints smoother samples as compared to other 3D Printers. The 3D volumetric structure for both the samples printed from different 3D Printers for the first printer setting is shown below in Figure 6.

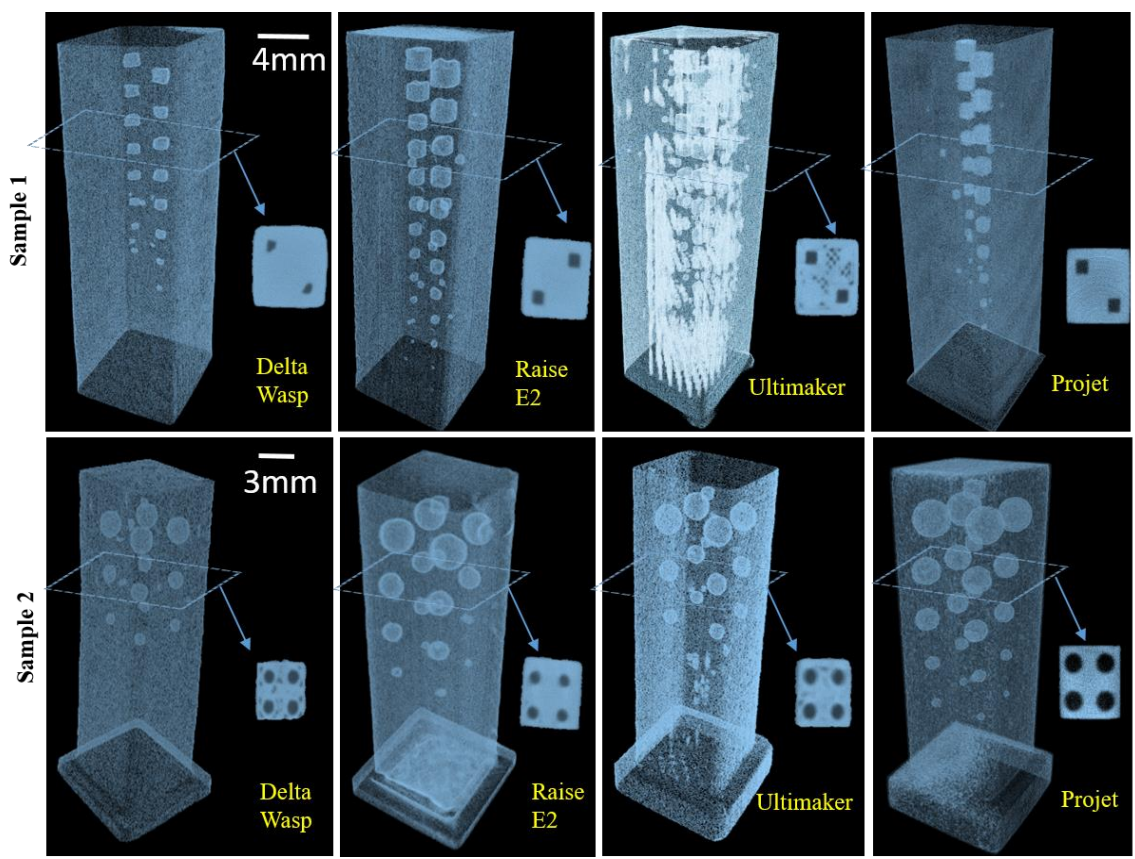

Figure 6: Volumetric surface and X-ray CT slice (at a certain height) of sample 1 and sample 2 printed from Delta Wasp, Raise, Ultimaker, and ProJet.

## 4.3. Qualitative Analysis of Surface Roughness

Volumetric and X-Ray CT slice analysis shows that there is a sufficient amount of roughness present in the samples fabricated by different 3D Printers. The accuracy of the different 3D Printers deviated from the actual performance. From the results of Volumetric and X-Ray CT analysis, it is observed that the amount of surface roughness present in samples printed by ProJet is least followed by Raise E2, Delta Wasp, and Ultimaker. That means ProJet is a good 3D printer, out of the four 3D printers that we analyzed.

The qualitative analysis of the surface roughness is done, to verify the results drawn from the different analysis. The region nearby to the cavities/voids were excluded for estimation of this parameter. Qualitatively, the internal volumetric surface of voids present in the samples is visualized in 3D by the volume rendering from the reconstructed X-Ray CT data. Figure 7.1 (a) and 7.1 (b) show the surface and internal volume distribution of sample 1 printed from Delta Wasp at the best optimal setting (first printer setting). Figure 7.1(c) shows the top view, void distribution, and boundaries of sample 1 printed from Delta Wasp. Figure 7.1 (d) illustrates the visualization of voids printed by the Delta wasp in sample 1. It is observed that the voids are not perfect cubical. There is distortion in the printed geometry of the void. Secondly, pores also existed in the printed sample due to an error in the printing nozzle. That is, the material is not deposited fully by the nozzle as it moved from one point to another and one layer to another layer. These defects are present in the sample in the form of porosity. All these defects together contribute to the surface roughness accumulated in the 3D printed sample.

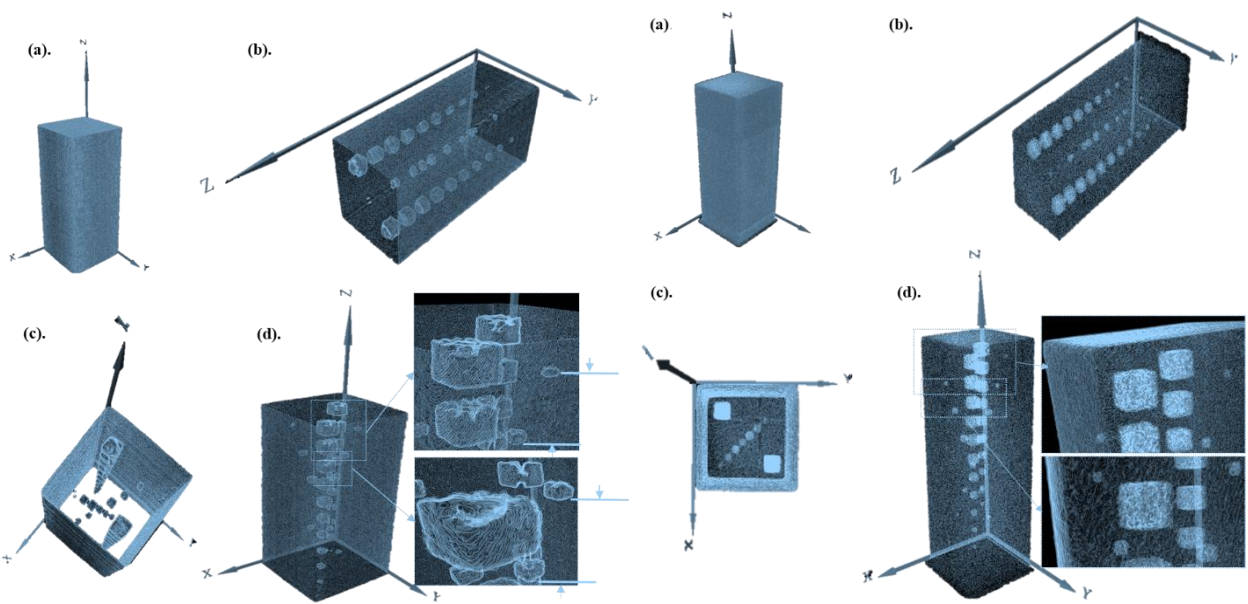

**Figure 7.1**                    **Figure 7.2**

Figure 7.1: (a), (b): CT image of sample 1 printed by Delta Wasp. (c). Top view of sample 1. (d): Surface roughness of voids in Sample 1 printed by Delta Wasp.

Figure 7.2: (a), (b): CT image of sample 1 printed by ProJet. (c). Top view of sample 1. (d): Surface roughness of voids in Sample 1 printed by ProJet.

Figures 7.2 (a) and 7.2 (b) show the surface and internal volume distribution of sample 1 printed from ProJet. Figure 7.2 (c) shows the top view of the void distribution and the sample surface. Figure 7.2 (d) illustrates the visualization of void distribution printed by ProJet in sample 1. It is observed that the surface of voids printed by ProJet is cubical and has no distortion in the printed geometry. Therefore, it is concluded that ProJet is a good 3D printer out of Delta wasp, Raise E2, Ultimaker Extended 2+, and ProJet. The results of qualitative analysis hold a good agreement with the Volumetric and X-Ray CT slice analysis.

### 4.4. X-Ray CT of Stone

The surface roughness of these 3D Printed samples is compared with sandstone pebble (hand-picked from Gangetic plane). We note that a typical sandstone pebble is formed by sand/sediment, deposited, and compacted underwater, which then become cemented to form a rock. This is so because the process of deposition of calcium and minerals in the sandstone pebble also occurs layer by layer and seems to be like the process of 3D Printing. It is found that the surface of this pebble is smoother than the surface of the samples printed from 3D printers. It means that nature may have a better 3D printing process than man-made 3D printers. The X-Ray CT slice and Cusp surface image for the sandstone pebble are shown in Figure 8 below. We found that it has a negligible porosity value.

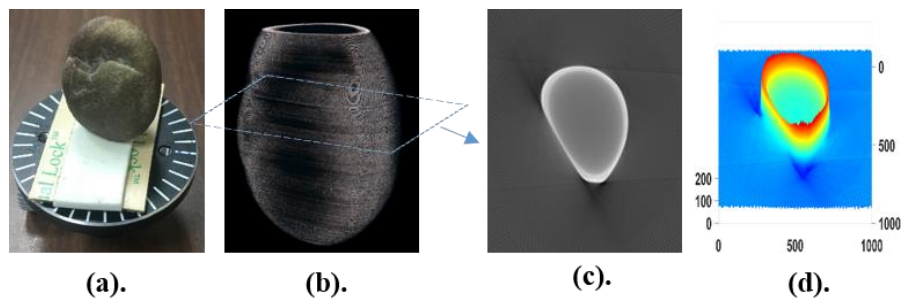

(a).       (b).       (c).       (d).

Figure 8: Sandstone Pebble: (a). X-Ray CT slice, (b). Cusp surface image, (c). Original image (d). 3D Surface Image.

## 4.5. X-Ray CT of Teeth

Further, the surface roughness of these 3D Printed samples and sandstone pebble is compared with the surface roughness of three deciduous teeth of a toddler (age 6 years, Indian female). The process of calcium deposition in teeth occurs by forming different layers of pulp, dentin, enamel, and gum. This can be correlated to the 3D printing process in which a sample is formed by depositing material layer by layer. Therefore, a comparison between the roughness accumulated in teeth because of calcium deposition and roughness in the samples fabricated from 3D Printing is made. It is also compared to the roughness of Sandstone pebble formed by calcium deposition in the river. The 3D analysis of the volume of the teeth is done by utilizing X-Ray CT and analyzing the reconstructed data from X-Ray CT. The 3D volume for the teeth data is shown in Figure 9. The fake colors are used to show the 3D depth. The light green color texture shows the inner volume, and the light blue color shows the surface of the teeth. We found these teeth sample also has a negligible porosity value.

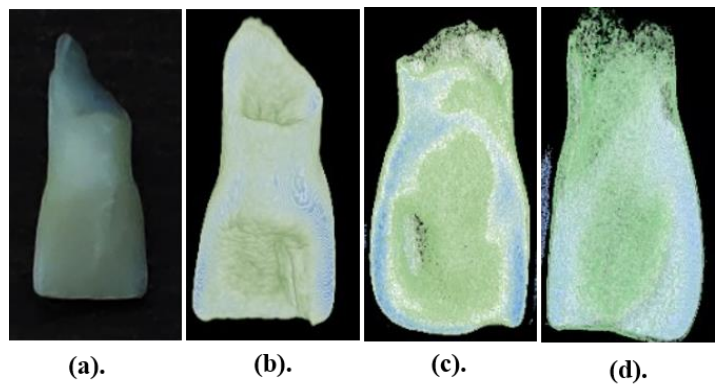

(a).       (b).       (c).       (d).

Figure 9: 3D volume of Teeth (a). Teeth image (b). Teeth 1 (c). Teeth 2 (d). Teeth 3.

## 4.6. Quantitative Analysis of Surface Roughness

To determine the surface roughness, cusp density and roughness of the samples are determined. The minimum cusp density and roughness found are plotted for all the printer settings for all the printers. Both the parameters are determined for the XY surface and XZ surface of CT slices separately. The plotted Cusp density and surface roughness for both the XY and XZ surfaces of CT slices are shown in Figure 10 below.

For sample 1 printed from Delta wasp, minimum cusp density for XY surface is found for the third printer setting, and for XZ surface, it is so for the sixth printer setting. For sample 2 printed from Delta wasp, minimum cusp density for XY surface is found for the fifth printer setting and the first printer setting is the minimum cusp density setting for XZ surface. For samples printed from Raise, minimum cusp density is found for the first printer setting for both the samples for both XY and XZ surfaces. For ProJet, minimum cusp density is found for sample 1 for both XY and XZ surfaces. The first printer setting is the minimum cusp density setting for both the samples printed from Ultimaker and for both the XY and XZ surfaces.

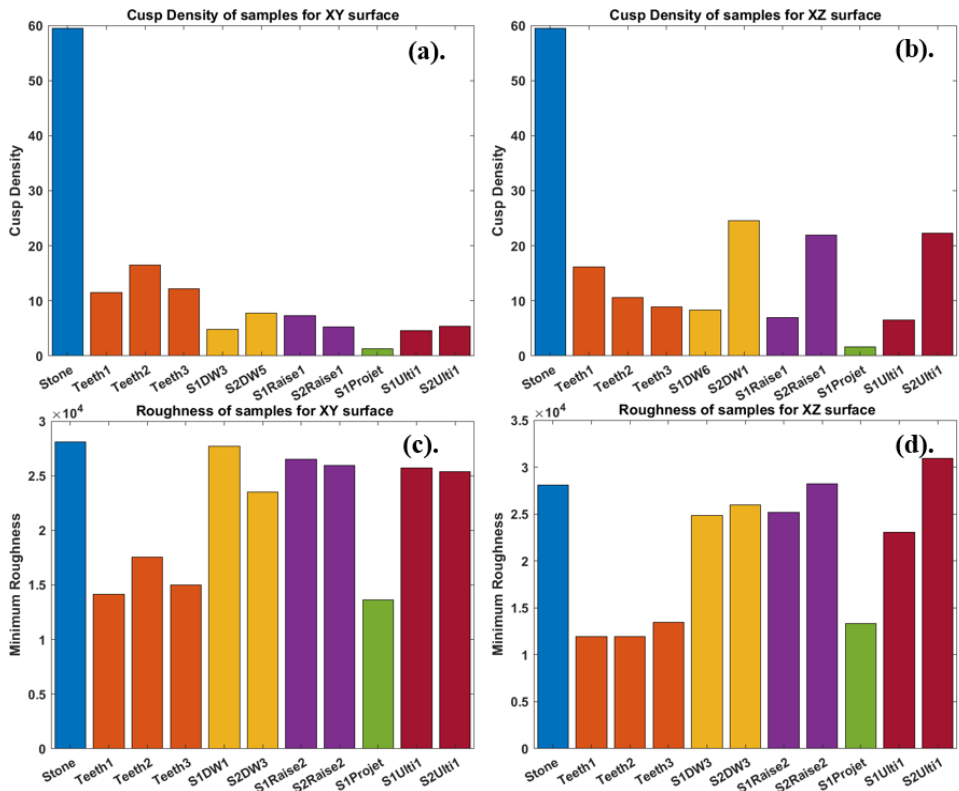

Figure 10: (a). Minimum Cusp Density of Samples for XY surface, (b). Minimum Cusp Density of Samples fo XZ surface, (c). Minimum roughness of Samples for XY surface, (d). Minimum roughness of Samples for XZ surface

For roughness measurement, minimum roughness for XY surface for sample 1 printed from the Delta is found at the first printer setting. For the XZ surface, the third printer setting is the minimum roughness setting. For sample 2 printed from Delta Wasp, the third printer setting gives minimum roughness for both XY and XZ surfaces. For samples printed from Raise, minimum roughness is found for the second printer setting for both the samples and for both the surfaces. For ProJet, minimum roughness is found in sample 1 for both XY and XZ surface analysis. The first printer setting is the minimum roughness setting for both the samples printed from Ultimaker Extended 2+.

**Table 2: Printer setting for minimum Cusp Density for XY surface**

| 3D Printer | Sample 1 | | Sample 2 | |
|---|---|---|---|---|
| | Printer setting | Cusp Density | Printer Setting | Cusp Density |
| **Delta Wasp** | Third | 4.8496 | Fifth | 7.7462 |
| **Raise E2** | First | 7.2638 | First | 5.2336 |
| **ProJet** | XHD mode | 1.3276 | XHD mode | 1.9423 |
| **Ultimaker** | First | 4.5611 | First | 5.4307 |

**Table 3: Printer setting for minimum Cusp Density for XZ surface**

| 3D Printer | Sample 1 | | Sample 2 | |
|---|---|---|---|---|
| | Printer setting | Cusp Density | Printer Setting | Cusp Density |
| **Delta Wasp** | Sixth | 8.3522 | First | 24.5121 |
| **Raise E2** | First | 6.9346 | First | 21.9980 |
| **ProJet** | XHD mode | 1.6586 | XHD mode | 4.7914 |
| **Ultimaker** | First | 6.5223 | First | 22.2990 |

**Table 4: Printer setting for minimum surface roughness for XY surface**

| 3D Printer | Sample 1 | | Sample 2 | |
|---|---|---|---|---|
| | **Printer setting** | **Roughness** | **Printer Setting** | **Roughness** |
| **Delta Wasp** | First | 2.7698e+04 | Third | 2.3506e+04 |
| **Raise E2** | Second | 2.6485e+04 | Second | 2.5944e+04 |
| **ProJet** | XHD mode | 1.3612e+04 | XHD mode | 4.0394e+04 |
| **Ultimaker** | First | 2.5694e+04 | First | 2.5363e+04 |

**Table 5: Printer setting for minimum surface roughness for XZ surface**

| 3D Printer | Sample 1 | | Sample 2 | |
|---|---|---|---|---|
| | **Printer setting** | **Roughness** | **Printer Setting** | **Roughness** |
| **Delta Wasp** | Third | 2.4820e+04 | Third | 2.5942e+04 |
| **Raise E2** | Second | 2.5175e+04 | Second | 2.8249e+04 |
| **ProJet** | XHD mode | 1.3359e+04 | XHD mode | 3.6834e+04 |
| **Ultimaker** | First | 2.3046e+04 | First | 3.0913e+04 |

## 4.7. Porosity Measurement and Optimal Printer Setting

The porosity measurement analysis is done to check the best printer and optimal printer setting. The porosity is measured for all the four 3D Printers and for all the printer settings. The inserted porosity in the CAD of samples 1 and 2 are 14.5 % and 14.8% respectively. Among the Delta Wasp, Raise E2, Ultimaker, and ProJet, maximum porosity (35.20%) is found in the samples printed from the Ultimaker, and the least porosity (14.56%) is found in the samples printed from ProJet. The bar graph is plotted for the minimum porosity and respective printer setting for both the samples and for all the 3D Printers. For sample 1 printed from Delta wasp, Raise E2, and Ultimaker, fifth, first, and first are the printer setting with 15.39 %, 21.20 %, and 35.20 % amount of porosity. For sample 2 printed from Delta Wasp, Raise E2, and Ultimaker, the third, second, and first are the printer setting with 17.19 %, 17.08 %, and 27.53 5 amount of porosity. The measured porosity is found to be higher than the inserted amount of porosity. The reason is that the printer's nozzle does not deposit material fully as it moves from one point to another and one layer to another layer. So, this concludes that a 3D printer prints the sample with a certain amount of porosity present in it. The porosity analysis is shown graphically in Figure 11 given below.

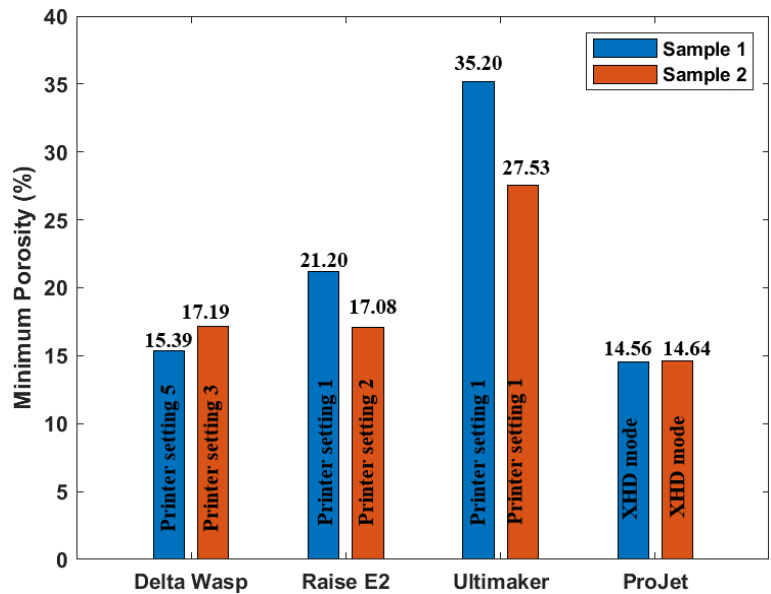

Figure 11: Porosity Measurement for different printers and printer settings

## 5. Conclusion

The purposed methodology provides a method to determine the surface roughness of 3D printed samples using X-Ray CT, both qualitatively and quantitatively. This method also gives an idea to check the quality and resolution of different 3D Printers and hence optimal printer settings. The conclusions drawn from the study are the following: -

1. Out of Ultimaker, Delta Wasp, Raise, and ProJet, minimum cusp density and roughness are found for ProJet. Therefore, ProJet is a good printer.
2. The porosity measurement also shows that minimum porosity in the sample printed from ProJet and so a good 3D printer out of the four.
3. Since the surface of voids present in the sample is smooth for ProJet, so samples printed from ProJet can withstand high strength. Therefore, when it is required that the sample can withstand high stress and strain, ProJet can be preferred to fabricate the samples.
4. Out of various printer settings for which we printed the samples, it is found that 50-micron layer height, 30mm/s nozzle speed, 100% infill density are the optimal printer setting for printing good resolution samples.
5. Surface roughness of teeth, sandstone pebble, and one of the best 3D printers (that is available to us) are compared. The amount of roughness present in teeth is found least. Amazingly, 3D printed Sample has better smoothness than sandstone pebble.
6. Results show that the biological process of teeth development is better than the process we have in 3D printers.

We recommend keeping minimum available layer height, minimum available nozzle speed and 100% infill density for best quality of 3D printing.

### Acknowledgments:

SK is thankful to IITR Institute Assistantship. MG is thankful to Tinkering Lab IIT Roorkee Staff.

### CRediT authorship contribution statement:

SK: Data Acquisition, Investigation, Writing, Visualization; AD: Data Acquisition; MG: Methodology, Data Acquisition, Investigation, Software, Writing, Visualization.

### Conflicts of Interest: The authors declare no conflict of interest.